\documentclass[aps,twocolumn,groupedaddress,showpacs,amssymb,prb,floatfix,preprintnumbers]{revtex4-1}
\usepackage{graphicx,color}
\usepackage{hyperref}
\usepackage{graphicx}             
\usepackage{amsmath}              
\usepackage{multirow}
\usepackage{xcolor,colortbl}

\newcommand{\as}         {$^{75}$As}

\newcommand{\LaOFeAs}    {${\mathrm{La}} {\mathrm{Fe}} {\mathrm{As}} {\mathrm{O}}$}

\newcommand{\LaOFFeAsx}  {${\mathrm{La}} {\mathrm{Fe}} {\mathrm{As}} {\mathrm{O}}_{1-x} {\mathrm{F}}_{x}$}
\newcommand{\SmOFFeAsx}  {${\mathrm{Sm}} {\mathrm{Fe}} {\mathrm{As}} {\mathrm{O}}_{1-x} {\mathrm{F}}_{x}$}
\newcommand{\REOFFeAsx}  {${\mathrm{RE}} {\mathrm{Fe}} {\mathrm{As}}{\mathrm{O}}_{1-x} {\mathrm{F}}_{x} $}

\newcommand{\BFCoAx}     {${\mathrm{Ba}} {\mathrm{Fe}}_{2-x} {\mathrm{Co}}_{x} {\mathrm{As}}_{2}$}
\newcommand{\BFRuAx}     {${\mathrm{Ba}} {\mathrm{Fe}}_{2-x} {\mathrm{Ru}}_{x} {\mathrm{As}}_{2}$}
\newcommand{\BFA}        {${\mathrm{Ba}} {\mathrm{Fe}}_{2} {\mathrm{As}}_{2}$}

\newcommand{\TN}      {$T_{N}$}
\newcommand{\TS}      {$T_{S}$}
\newcommand{\Tc}      {$T_{c}$}
\newcommand{\Tmax}    {$T_{\mathrm{max}}$}

\newcommand{\slrrt}     {$(T_1T)^{-1}$}
\newcommand{\slrr}      {$T_1^{-1}$}
\newcommand{\slrrab}      {$T_{1ab}^{-1}$}
\newcommand{\slrrc}      {$T_{1c}^{-1}$}
\newcommand{\ssrr}      {$T_2^{-1}$}
\newcommand{\ratio}     {$R=T_{1ab}^{-1}/T_{1c}^{-1}$}

\newcommand{\mpoint}{\,.}					

\definecolor{orange}{RGB}{255,165,0}
\definecolor{violett}{rgb}{0.5,0.0,0.5}
\definecolor{kblue}{RGB}{0,0,128}
\definecolor{olive}{RGB}{0,139,0}

\begin{document}

\thispagestyle{myheadings}

\title{A unified phase diagram of F-doped LaFeAsO by means of NMR and NQR parameters}

\author{Hans-Joachim Grafe~$^{1}$, Piotr Lepucki~$^{1}$, Markus Witschel~$^{1,3}$, Adam P. Dioguardi~$^{1}$, Rhea Kappenberger~$^{1}$, Saicharan Aswartham~$^{1}$, Sabine Wurmehl~$^{1,2}$, Bernd B\"uchner~$^{1,2}$}
\affiliation{$^1$Leibniz IFW Dresden, Institut f\"ur Festk\"orperforschung, Helmholtzstraße 20, D-01069 Dresden, Germany\\ $^{2}$Institut f\"ur Festk\"orperphysik, Technische Universit\"at Dresden, D-01171 Dresden, Germany\\ $^{3}$Wests\"achsische Hochschule Zwickau, Fakult\"at Physikalische Technik/Informatik, PSF 201037, 08012 Zwickau, Germany}

\date{\today}

\begin{abstract}
We present \as\ Nuclear Magnetic and Quadrupole Resonance results (NMR, NQR) on a new set of \LaOFFeAsx\ polycrystalline samples. Improved synthesis conditions led to more homogenized samples with better control of the fluorine content. The structural$\equiv$nematic, magnetic, and superconducting transition temperatures have been determined by NMR spin-lattice relaxation rate and AC susceptibility measurements. The so-determined phase diagram deviates from the published one especially for low F-doping concentrations. However, if the doping level is determined from the NQR spectra, both phase diagrams can be reconciled. The absence of bulk coexistence of magnetism and superconductivity and a nanoscale separation into low-doping-like and high-doping-like regions have been confirmed. Additional frequency dependent intensity, spin-spin, and spin-lattice relaxation rate measurements on underdoped samples at the boundary of magnetism and superconductivity indicate that orthorhombicity and magnetism originate from the low-doping-like regions, and superconductivity develops at first in the high-doping-like regions.
\end{abstract}


\maketitle

\section{Introduction}

The transition from antiferromagnetism (AF) to superconductivity (SC) is one of the most studied issues in iron-based as well as in other unconventional superconductors such as cuprates and heavy fermions. Thereby, the accurate identification of structural, magnetic and superconducting phase boundaries and their transition temperatures is as important as the exact determination of the doping level. \cite{MartinelliCRPhys2016,PengDaiRevModPhys2015,KeimerNat2015, WirthNatMatRev2016} In iron pnictides, regarding the relation between magnetism and superconductivity, controversial results are still reported even within one family and depending on the dopant. Mutual exclusion between magnetic and superconducting phases is accompanied by microscopic phase coexistence (see Ref.~\onlinecite{MartinelliCRPhys2016} and references therein). For example, in \BFCoAx , the majority of experimental results point towards a microscopic coexistence of superconductivity and magnetism for underdoped samples with $x \leq 0.055$, though accompanied by slowly fluctuating inhomogeneous magnetic correlations. \cite{PrattPRL2009,BernhardPRB2012,LaplacePRB2009,JulienEPL2009,DioguardiPRL2013,DioguardiPRB2015} On the other hand, for \BFRuAx , a nanoscale texture of magnetism has been reported, i.e. an inhomogeneous electronic state at the nanoscale in real space that also leads to suppression of magnetism and appearance of superconductivity \cite{LaplacePRB2012}. Similarly, a nanoscale SC-AF hybrid state has been found in SrFe$_2$As$_2$ under high pressure, and has been taken as an example of a self-organized heterogeneous structure in a clean system \cite{KitagawaPRL2009}. In addition, for the hole doped side of the phase diagram, e.g. in RbFe$_2$As$_2$ and CsFe$_2$As$_2$, charge order seems to be well established \cite{CivardiPRL2016,MoroniPRB2019}, and can be explained by the closeness to selective Mottness at half-filling \cite{deMediciPRL2014}. 

In \REOFFeAsx\ (RE = rare earth), the situation is even more complicated due to the uncertain determination of the fluorine content. For magnetic rare earths, previous $\mu$SR results found 100\% magnetic volume fraction in underdoped, superconducting samples, i.e. ostensible coexistence of magnetism and superconductivity \cite{DrewNatMat2009}. On the other hand, for \LaOFFeAsx\ coexistence of static magnetism with a sizable moment and superconductivity appear to be excluded \cite{LuetkensNatMat2009}. Nevertheless, there are also reports of coexistence of magnetism and superconductivity in underdoped \LaOFFeAsx , albeit with tiny static magnetic moments on the order of 0.006~$\mu_B$, i.e. about 100 times less than in the parent LaFeAsO \cite{NakaiPRB2012,YangSciChin2018}. 

In a previous study, we demonstrated that a nanoscale separation exists in \REOFFeAsx\ dividing the samples into low-doping-like regions that fully account for orthorhombicity and magnetism, and high-doping-like regions that accommodate superconductivity \cite{LangPRB2016}. The spatial competition of these two ground states excludes the coexistence picture for \REOFFeAsx . Then, the 100\% magnetic volume fraction from $\mu$SR does not mean that 100\% of the sample is intrinsically magnetic, but that the field from the ordered regions can be felt by the muons in the superconducting regions due to their nanoscale size. Furthermore, from the particular pattern of the NQR spectra we found an indirect way of determining the doping level, which allowed us to reconstruct a phase diagram including samples from various groups prepared over a period of a decade. However, this approach could only provide a rough estimate of the absolute fluorine doping level. \cite{LangPRB2016,SannaPRB2010}

Here, we show results on a new set of samples where we paid particular attention to the control of the F doping concentration. In Sec. \ref{sec:NQRspec} we use the NQR spectra to determine the doping levels, which can be compared to previously published results \cite{LuetkensNatMat2009,LangPRB2016}. From the temperature dependence of the NQR frequency, $\nu_Q$, we infer the difference between the doping dependence of $\nu_Q$ and the changes of the quadrupole parameters $(\nu_a, \nu_b, \nu_c )$, at the nematic transition temperature, \TS . From the temperature dependence of the NQR intensity of underdoped samples, we can prove that magnetism originates in the low-doping-like regions, and that the observed loss of signal intensity is due to the appearance of static magnetic fields in part of the samples and not due to a dynamical wipeout. In Sec. \ref{sec:T1andsusce} we show the spin-lattice relaxation rate, \slrrt , measured by NMR and NQR, and the AC susceptibility. From these measurements, we could accurately determine the magnetic, structural, and superconducting transition temperatures \TN , \TS , and \Tc , respectively. We find no evidence for bulk coexistence of magnetism and superconductivity in \LaOFFeAsx . Finally, in Sec. \ref{sec:phasedia}, we construct a phase diagram and compare the new samples to previously published data for nominal as well as for NQR determined F doping levels.

\section{Experimental Details}

\subsection{Synthesis \& characterization}

Polycrystalline samples of F-doped LaFeAsO were prepared using solid-state reaction as described in Refs.~\onlinecite{AlfonsovPRB2011,KappenbergerJCrysGro2018,KappenbergerPRB2018}. First, the precursor LaAs was prepared using La (Chempur, 99.9~$\%$) and As lumps (Chempur, 99.999~$\%$) in a stoichiometric ratio. The reaction then takes place in an evacuated quartz tube placed in a two-zone furnace. For the second step, the resulting LaAs was mixed with Fe (Alfa Aesar, 99.998~$\%$), Fe$_2$O$_3$ (Chempur, 99.999~$\%$) and FeF$_3$ in the corresponding ratio that lead to the desired F concentration. Those starting materials were homogenized by grinding in a mortar and afterwards in a ball mill. The ball milling ensures better homogeneity and proper intermixing compared to previous samples which were ground in a mortar only. The resulting powders were pressed into pellets under Ar atmosphere using a force of 20 kN and subsequently annealed in an evacuated quartz tube. Clearly, the use of different presursors, the ball milling, and a longer annealing time have a significant impact on the samples quality especially regarding purity and homogeneity of the F-content.
 
To investigate the sample quality, several complimentary methods were used for characterization. Microstructure and composition were analyzed by scanning electron microscopy (SEM) on the as synthesized pellets. The analysis was carried out using a Zeiss EVOMA15 with AzTec software. The acceleration voltage was 30~kV. EDX spectra on 15-30 points or small areas on each sample were collected to confirm the elemental content. Powder x-ray diffraction was measured on powdered polycrystalline pellets and confirms the tetragonal symmetry. A structural model was obtained by Rietveld analysis using the FullProf software package \cite{RietveldJAppCrys1969,RoisnelMatSciFor2001}. The superconducting transition temperatures have been determined by SQUID magnetometry.

\subsection{NMR}
\label{sec:NMR}

The NMR spectra show the typical powder pattern of a polycrystalline sample, which is expected for a nucleus (\as , I =3/2) that is affected by the quadrupole interaction \cite{GrafePRL2008}. The spin-lattice relaxation rate \slrr\ has been measured at the high frequency peak of the NMR powder pattern which corresponds to the orientation $H || ab$ of the grains of the powder sample. It is therefore directly comparable to single crystal data measured with $H || ab$ \cite{FuPRL2012,OkPRB2018}. \slrr\ has been measured by inversion recovery, and the recovery of the nuclear magnetization was fitted to the relaxation formula for the central transition of a nuclear spin I = 3/2: $M_z(t) = M_0[1-f(0.9\exp(-(6t/T_1)^\beta) + 0.1\exp(-(t/T_1)^\beta))]$, where $M_z(t)$ and $M_0$ are the nuclear magnetization, $t$ the time, $f$ the inversion factor which is ideally 2. $\beta$ is a stretching exponent which indicates a distribution of spin-lattice relaxation times if $\beta < 1$, as has been observed before in \LaOFFeAsx \cite{HammerathPRB2013}.

The spin-lattice relaxation rate is determined by the fluctuating hyperfine fields (due to the electronic spins of the four nearest neighbor Fe sites of the As) at the nuclear site (As), and therefore measures the imaginary part of the dynamical spin susceptibility $\chi^{\prime\prime}$ of the electronic spin system at the NMR frequency, $\omega$: 
\begin{equation}
(T_1T)^{-1} \propto \sum_{\vec{q},\alpha,\beta}F_{\alpha,\beta}(\vec{q}) \frac{\chi^{\prime\prime}_{\alpha,\beta}(\vec{q},\omega)}{\omega} \, \mpoint
\label{eq:T1T}
\end{equation}
Here, $F_{\alpha,\beta}(\vec{q})$ denotes the hyperfine form factors with $\alpha,\beta = {x,y,z}$ which are determined by the Fourier transformation of the hyperfine coupling tensor. The form factors have been calculated \cite{SmeraldPRB2011,KissikovNatCom2018}, and due to the particular position of the As, in-plane stripe type spin fluctuations are enhanced, whereas out-of-plane fluctuations are reduced. This leads to a ratio of the spin-lattice relaxation rates for different field orientations of $R=\frac{T_1^{-1}(H||ab)}{T_1^{-1}(H||c)} = 1.5$ for isotropic spin fluctuations. When the spin fluctuations couple to the nematic order, they become anisotropic, and the ratio $R$ is strongly enhanced. \cite{KitagawaJPSJ2009,SKitagawaPRB2010,NakaiPRB2012} This has been observed in all iron pnictide families that exhibit strong spin fluctuations \cite{DioguardiPRB2015,KissikovNatCom2018,GrafePRB2014,ZhouPRB2016,NingPRB2014}.

\subsection{NQR}

The NQR spectra have been measured point by point with a phase-coherent Tecmag NMR spectrometer. The sample probe has been tuned automatically to the desired frequency, and the intensity has been integrated at each frequency step \cite{YannicPhD2016}. The quadrupole frequency $\nu_Q$ is determined by the electric field gradient (EFG) at the nucleus:

\begin{align}
\nu_{Q} = & \frac{eQV_{zz}}{2h} \cdot \sqrt{1+\frac{\eta^2}{3}},
\label{eq:nuNQR}
\end{align}

where the asymmetry parameter $\eta = \frac{V_{xx}-V_{yy}}{V_{zz}}$, and $V_{\alpha \alpha}$ are the diagonal components of the EFG tensor. In the tetragonal phase of iron pnictides, the components in the $ab$ plane are equal and smaller than the component along $c$: $|V_{xx}| = |V_{yy}| < |V_{zz}|$, i.e. $\eta = 0$. For NMR on single crystalline samples with $(x,y,z) = (a,b,c)$, each component of the EFG, i.e. $\nu_a$, $\nu_b$, and $\nu_c$ can be determined independently by orienting the external magnetic field along the crystallographic directions $a,b,c$. Below the structural transition, $\eta$ becomes larger than 0 and $\nu_{c}$ is not necessarily the largest component anymore. Note that this large change of $\eta$ was one of the first indications that the structural transition involves electronic degrees of freedom, i.e. the nematic transition is electronically driven and not just by the structural effects or structural misfits which grow with decreasing temperature. \cite{KitagawaJPSJ2008} Therefore, the term nematic transition has been established \cite{FernandesNatPhys2014}. 

The spin-lattice relaxation rate was also measured by NQR. Here, since the largest component of the EFG is parallel to the crystallographic $c$-axis, \slrr\ corresponds to \slrr\ measured by NMR in a magnetic field with the orientation $H || c$. The only difference to \slrr\ as measured via NMR is the applied magnetic field, and therefore also the applied frequency. However, an advantage is that one can measure \slrr\ on the two different NQR peaks which occur for underdoped samples, and thereby gain more information on the nature of these different peaks \cite{OkaPRL2012}. For \slrr\ measured by NQR, the recovery function is  $M_z(t) = M_0[1-f(\exp(-(3t/T_1)^\beta)$.

The spin-spin relaxation rate, \ssrr , has been measured by a Hahn Echo pulse sequence in order to correct for the loss of signal intensity at low temperature, and to investigate the origin of this loss of intensity. The relaxation of the nuclear magnetization in the $xy$ plane has been fitted to a combined exponential and Gaussian form: 

\begin{equation}
M_{xy}(t) = M_0 \cdot \exp(-(2t/T_{2})^\beta) \cdot \exp(-(2t)^2/2T_{2G}^2) \, 
\label{eq:T2}
\end{equation}

where $M_0$ is the nuclear magnetization, $t$ the time between the $90^\circ$ and $180^\circ$ pulses. $T_2$ is the exponential spin-spin relaxation time, and $T_{2G}$ the Gaussian relaxation time. Such a form has been used to fit the $T_2$ data in other iron pnictides, too \cite{BossoniPRB2013,DioguardiPRB2015}. $\beta \leq 1$ accounts for a distribution of spin-spin relaxation times similar to the distribution of the spin-lattice relaxation times.

\section{NQR spectra}
\label{sec:NQRspec}

\begin{figure}
\begin{center}
 \includegraphics[width=\columnwidth,clip]{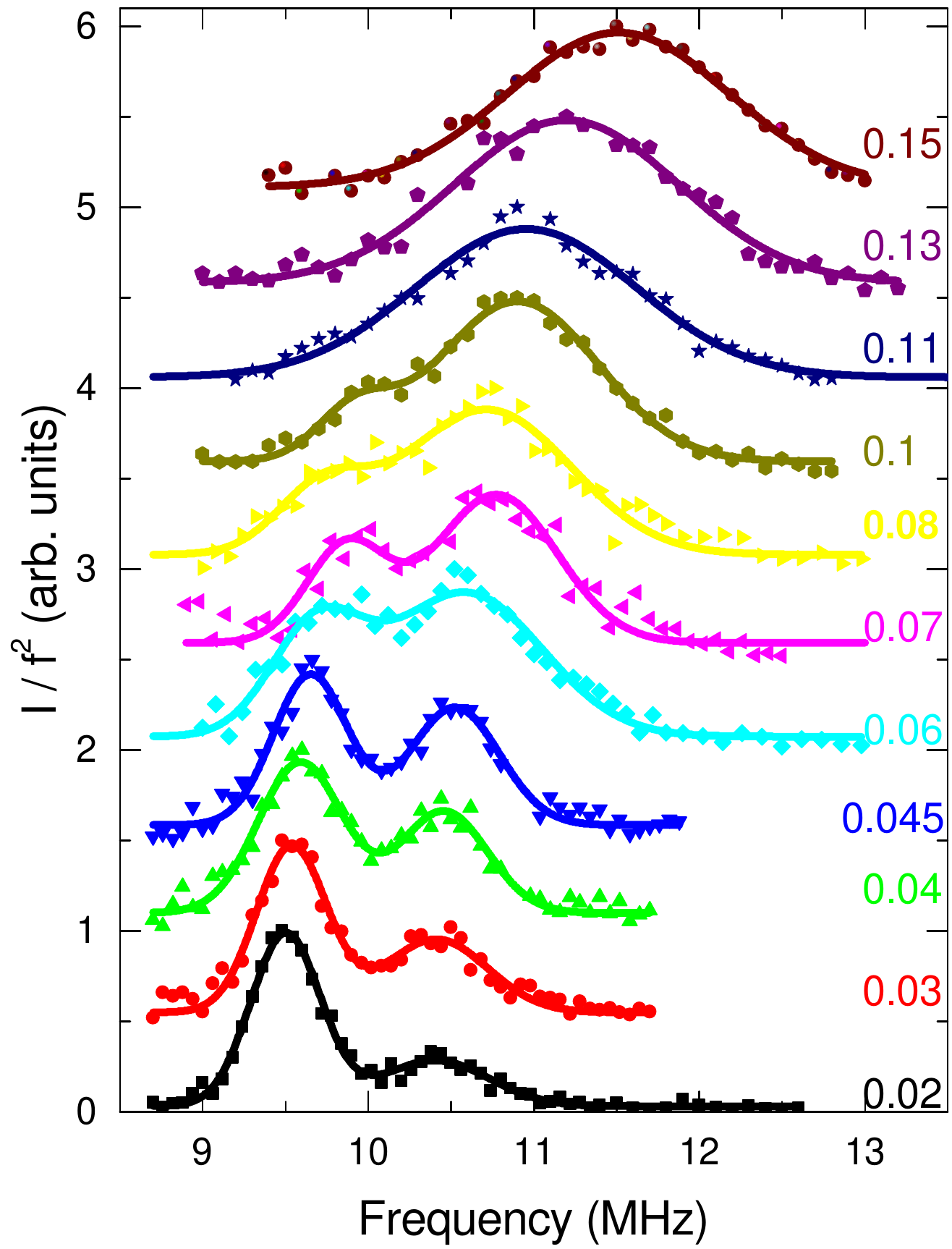}
 \caption{(Color online) The NQR spectra of \LaOFFeAsx\ for different doping levels $x$ at 300 K (except for $x=0$). The solid line represents a double Gaussian fit for underdoped samples ($x \leq 0.1$) and a single Gaussian fit for higher doping levels ($x > 0.1$). The area of the high frequency Gaussian peak divided by the whole intensity of the spectrum is the high frequency weight $w^H$ (see also \cite{LangPRB2016}).}
\label{fig:NQRspec}
\end{center}
\end{figure}

The NQR spectra for the different nominal doping levels at 300~K are shown in Fig.~\ref{fig:NQRspec}. The doping dependence of the \LaOFFeAsx\ NQR spectra is already well known \cite{LangPRL2010,LangPRB2016}. Specifically, the spectrum of the undoped compound has a single narrow peak at a frequency of 9.5~MHz (not included in Fig.~\ref{fig:NQRspec}, but in Refs.~\onlinecite{LangPRL2010,LangPRB2016}). The underdoped samples (up to roughly x~=~0.1) display an additional peak at $\sim$10.5~MHz. The spectral weight $w^H$, i.e. the integrated spectral area, and frequency of this peak increase with F doping. Finally, the spectra of the optimally- and overdoped samples (with $x > 0.1$) have a single broad peak whose frequency increases with F doping. As will be seen below, the samples with $x=0.02$ and $x=0.03$ undergo a structural transition and order long-range magnetically. The sample with $x=0.04$ does not show a structural transition, and also no long range magnetic order. On the other hand, superconductivity seems to be only very weak in this sample, therefore, this sample is considered to be at the transition from bulk magnetic order to superconductivity. All samples with a nominal $x \geq 0.045$ are superconducting and do not show indications of structural or long range magnetic order. 

The quadrupole frequency strongly depends on doping, as has been found in many other iron pnictides \cite{NingPRB2014,BaekPRB2011,HiranoJPSJ2012,BaekEPJB2012}. In Ref.~\onlinecite{LangPRB2016}, we have shown that the additional doping-dependent peak in the NQR spectrum is due to a nanoscale separation into low-doping-like and high-doping-like regions, and not due to a static effect of the F dopants as discussed in Ref.~\onlinecite{YangSciChin2018}. This has been confirmed by other groups, where it was found that the spectral weight of the two peaks also changes with other dopants such as Ru, Mn, or Co \cite{SannaPRL2011,MoroniPRB2017,Lepuckiunpub}. The increase of the spectral weight of the high frequency peak $w^H$ with increasing F concentration can then be used to construct a phase diagram which is indicative of the true F concentration. Here, it will be used to compare the doping levels of the new samples with old ones.

\subsection{Temperature dependence of the NQR spectra}

\begin{figure}
\begin{center}
 \includegraphics[width=\columnwidth,clip]{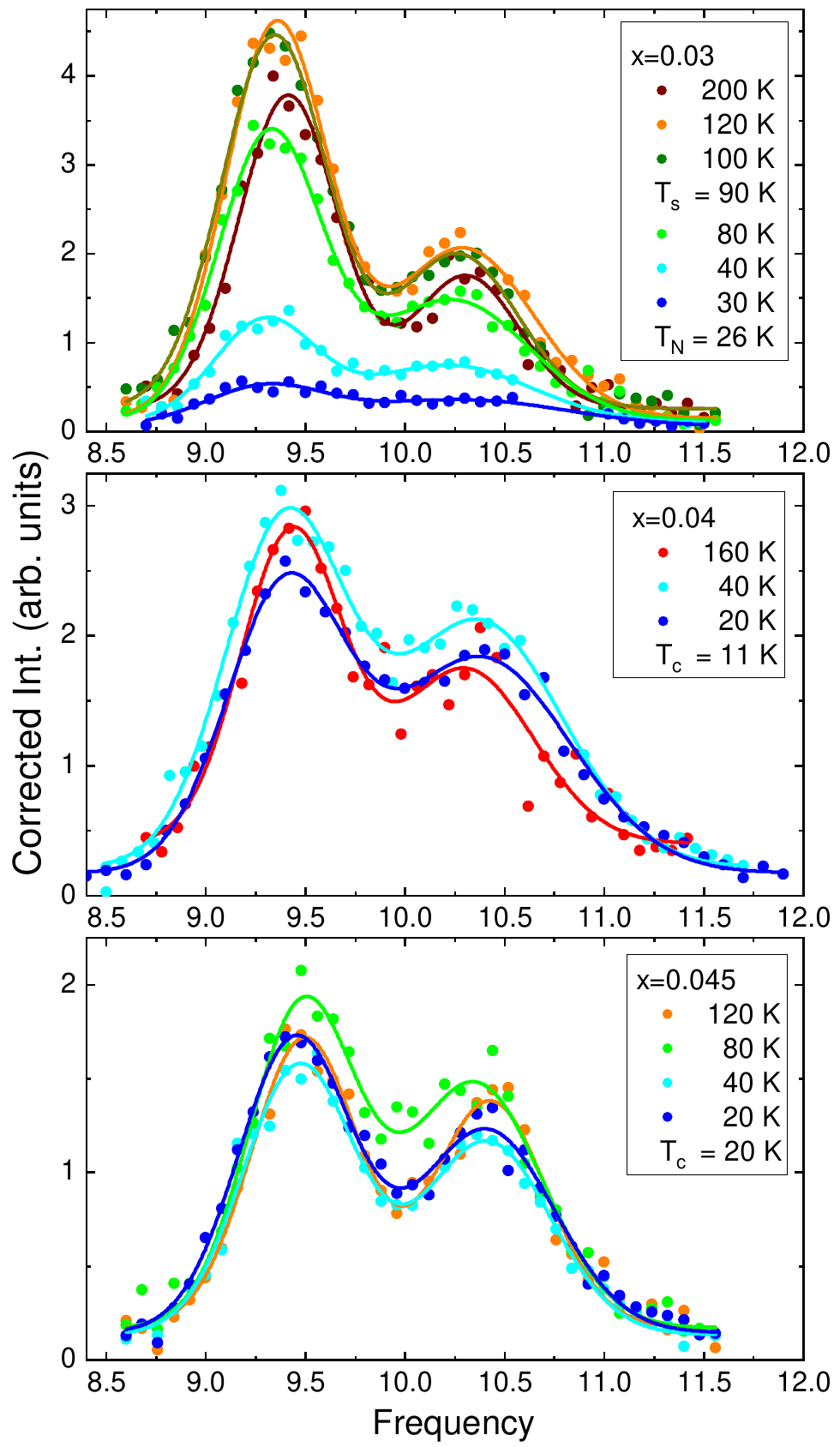}
 \caption{(Color online) The temperature dependence of the NQR spectra for $x=0.03, 0.04$, and $0.045$. For the magnetic sample with $x=0.03$, the total intensity decreases already below $T_S$. For the other doping levels, the intensity is constant. Solid lines are fits of the spectra to two Gaussian lines.}
\label{fig:TabhNQR}
\end{center}
\end{figure}

\begin{figure}
\begin{center}
 \includegraphics[width=\columnwidth,clip]{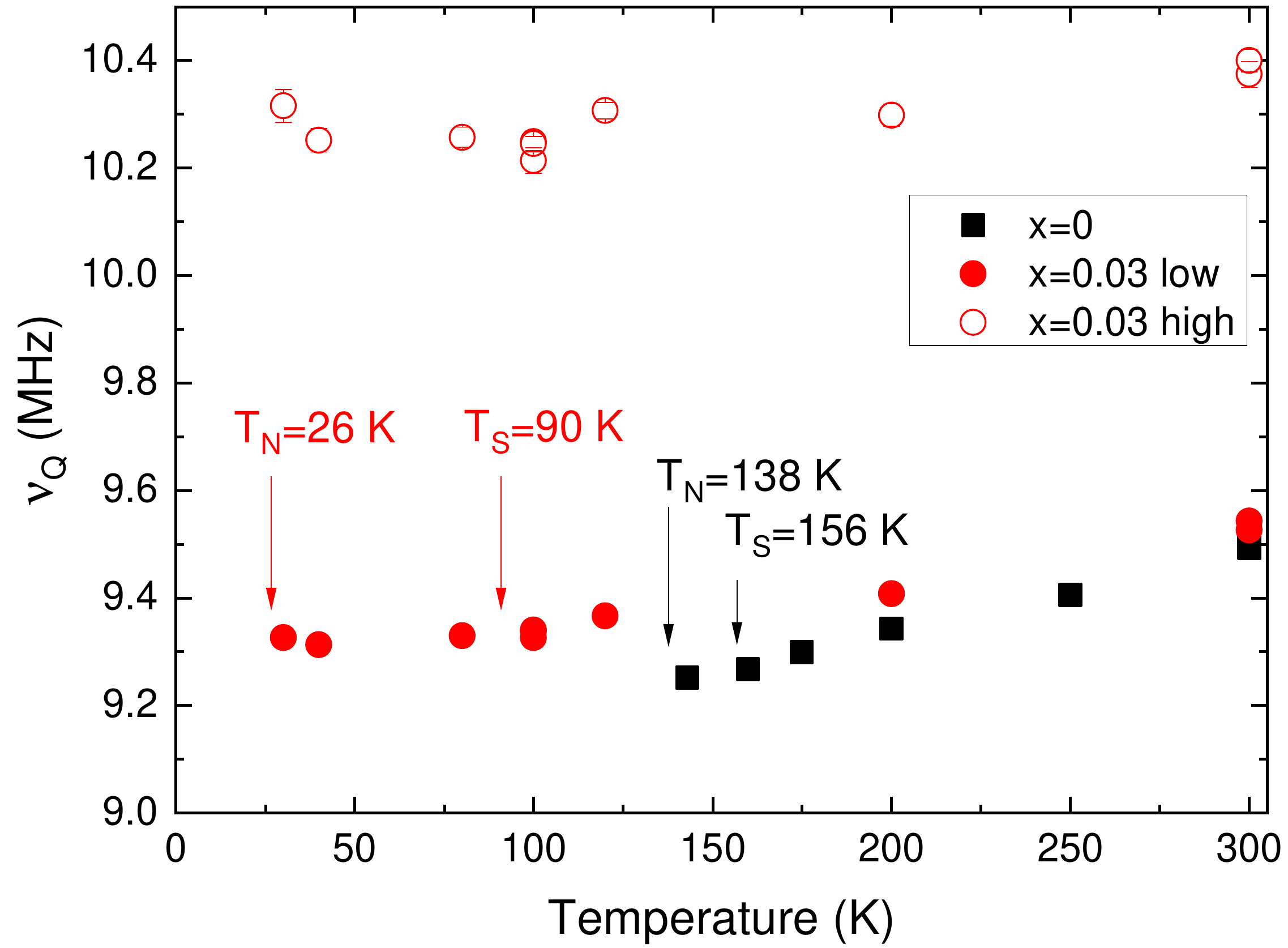}
 \caption{(Color online) The temperature dependence of the NQR frequency, $\nu_Q$, for \LaOFFeAsx\ with $x=0$ (black squares, from \cite{LangPRB2016}	) and $x=0.03$ (red dots).}
\label{fig:nuQ}
\end{center}
\end{figure}

In Fig.~\ref{fig:TabhNQR} we show temperature dependent NQR spectra of the samples with $x=0.03$ (last sample that orders magnetically), $x=0.04$ (intermediate sample between magnetism and superconductivity), and $x=0.045$ (superconducting sample, no magnetism). The intensities of the spectra have been corrected by the number of scans, the Boltzmann factor, and by the square of the frequency. A correction for the spin-spin relaxation time $T_2$ is not necessary, as will be shown below. We extract the quadrupole frequencies from the maxima of the peaks. $\nu_Q$ decreases weakly with decreasing temperature, but does not show a marked change at the nematic transition $T_S$ (see Fig.~\ref{fig:nuQ}). This is surprising, because in all iron pnictide compounds, the quadrupole parameters $\nu_{a,b,c}$ which determine $\nu_Q$ (see Eqn. \ref{eq:nuNQR}), change abruptly at the nematic transition, \TS \cite{ZhouPRB2016,KitagawaJPSJ2008,FuPRL2012,KitagawaJPSJ2009,BaekPRB2009}. 

For LaFeAsO, the change of the quadrupole parameters and also of the orthorhombic distortion, $\delta$, are not as large as in \BFA , nevertheless $\nu_{a,b,c}$ and $\delta$ change substantially below $T_S$. For example, $\nu_a$ increases from 4.6~MHz to 5.2~MHz at 145~K, and $\nu_b$ and $\nu_c$ drop from 4.6~MHz to 3.9~MHz and from 9.2~MHz to 8.8~MHz, respectively. \cite{WangJMMM2019,FuPRL2012} Despite the marked change of $\nu_{a,b,c}$ at the nematic transition and the strong doping dependence of $\nu_Q$, there is no change of $\nu_{Q}$ below \TS\ within the experimental errors. This is because the changes of $\nu_c$ and $\eta$ act in the opposite direction on $\nu_Q$ (compare Eqn. \ref{eq:nuNQR} and Ref.~\onlinecite{MoroniPRB2019}), and is therefore evidence for an orbital order, i.e. a rearrangement of charges in the relevant orbitals. On the other hand, the doping dependence of the quadrupole frequency indicates an increase of the number of charges in the Fe orbitals which are hybridized with the As $p$ orbitals. 

\subsection{Temperature dependence of the intensity and spectral weight}

\begin{figure}
\begin{center}
 \includegraphics[width=\columnwidth,clip]{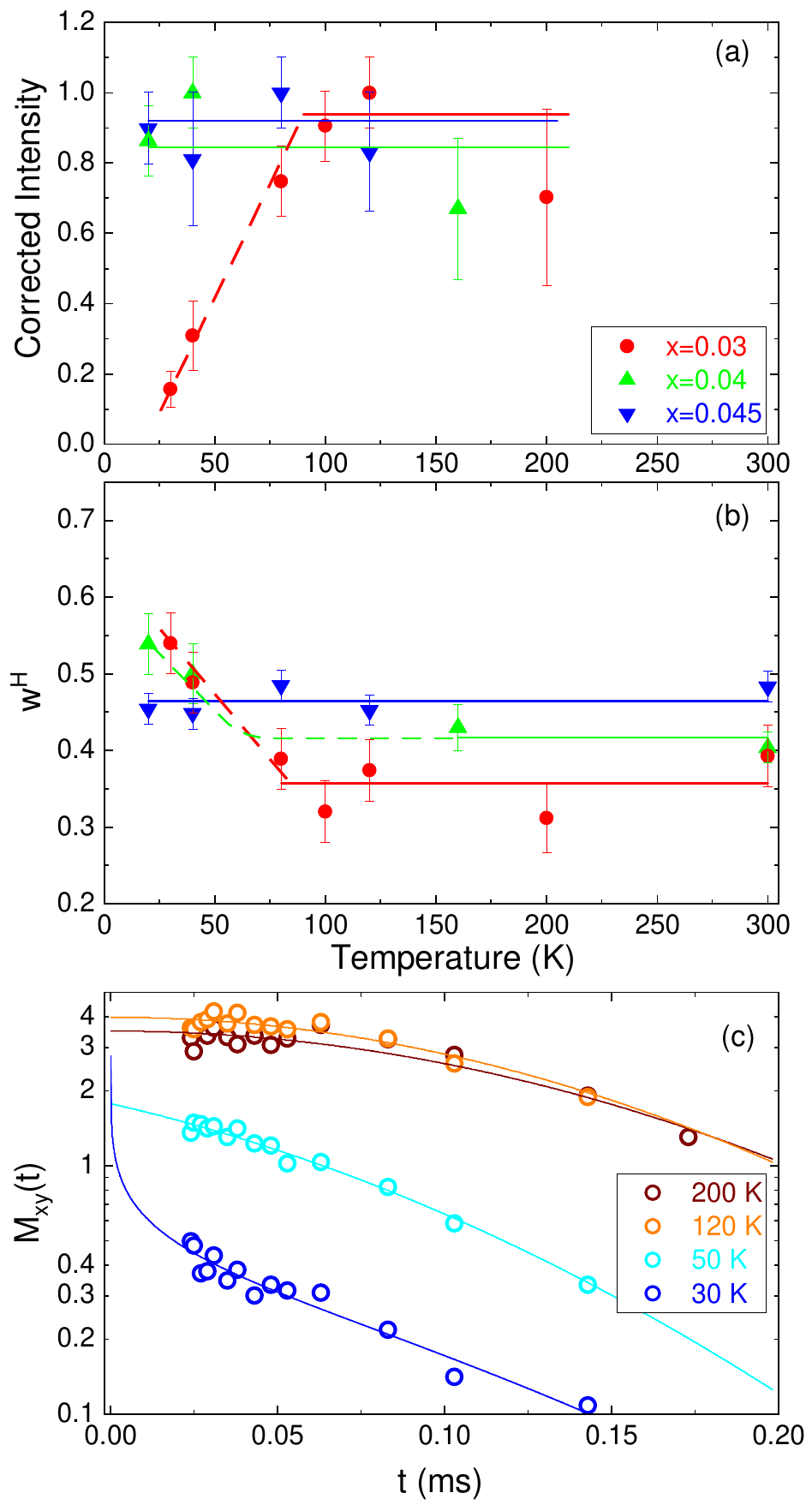}
 \caption{(Color online) (a) The temperature dependence of the corrected total intensity of the NQR spectra for $x=0.03, 0.04$, and $0.045$. (b) The temperature dependence of the spectral weight of the high frequency peak, $w^H$. For $x=0.03$, $w^H$ increases already below $T_S$. Solid lines are linear fits, dashed lines are guides to the eye. (c) The temperature dependent echo decay curves for $x=0.03$, corrected in intensity, by number of scans, and Boltzmann factor. The lines are fits to Eqn.~\ref{eq:T2}, and are extrapolated to 0~ms.}
\label{fig:Tabhparam}
\end{center}
\end{figure}

 The intensity of the sample with $x=0.03$ starts to decrease at about 80 K, which is just below the structural transition temperature $T_S = 90$ K, but far above the bulk magnetic ordering temperature $T_N = 26$ K (see Fig.~\ref{fig:TabhNQR} and Fig.~\ref{fig:Tabhparam} (a)). The decrease of the intensity for $x=0.03$ can be either due to static magnetic fields which shift or broaden the resonance substantially, or due to a dynamical wipeout effect, which occurs e.g. in underdoped \BFCoAx\ due to fast spin-spin and/or spin-lattice relaxation rates \cite{DioguardiPRL2013,DioguardiPRB2015}. The dynamical wipeout effect is best known from cuprates, where it has been shown that it is indeed a dynamical effect \cite{CurroPRL2000,JulienPRB2001}, and that the total intensity can be recovered if the time between the pulses of a Hahn echo sequence is reduced to a minimum \cite{PelcPRB2017}. In contrast to $x=0.03$ and to \BFCoAx , the intensities of the underdoped superconducting samples with $x=0.04$ and $x=0.045$ do not change strongly down to $T_c$. Below $T_c$, the intensity changes due to limited penetration depth of the rf pulses (not shown).

In order to gain more insight into the origin of the intensity loss, we plot in Fig.~\ref{fig:Tabhparam} (b) the spectral weight of the high frequency peak, $w^H$. Similar to the total intensity of the spectra in Fig.~\ref{fig:Tabhparam} (a), $w^H$ changes below $T_S$: the intensity of the low frequency peak decreases more strongly than that of the high frequency peak. This is already visible in the spectra at 30~K and 40~K in Fig.~\ref{fig:TabhNQR}, and indicates that magnetism develops at first in the underdoped region, and reduces the spectral weight of the low frequency peak by internal fields and/or fast relaxation. On the other hand, the high frequency peak seems to be less affected by these static and/or fluctuating internal fields. For $x=0.04$, the high frequency spectral weight $w^H$ seems to increase somewhat at low temperatures, too. However, the total intensity of this sample does not seem to change. Note that the spectra at 300 K have not been included into the $T$ dependence of the total intensities, because they have been measured with a resonant circuit with an optimized Q factor in order to maximize the signal intensity. For the $T$ dependent spectra, a $T$ independent resistor has been added to the resonant circuit in order to reduce effects of a temperature dependent $Q$ factor on the signal intensity. 

In order to find out whether the loss of signal intensity is a static or a dynamical effect, we did spin-spin relaxation rate measurements on the low-frequency peak. The decay of the nuclear magnetization $M_{xy}$ versus the separation time of the two pulses of the Hahn echo sequence is shown in Fig.~\ref{fig:Tabhparam} (c) for several temperatures. The shortest possible separation time was 23~$\mu$s. The decay of $M_{xy}$ has been fitted to Eqn.~\ref{eq:T2}, and extrapolated to $t=0$~ms. For 50~K, the fit leads to a stretching exponent of $\beta = 1$, i.e. no distribution of spin-spin relaxation times, and the nuclear magnetization $M_{xy}$ does not fully recover to its value at high temperatures, meaning that a fraction of the nuclei should feel static hyperfine fields which shift or broaden the spectra substantially and lead to the loss of signal intensity. For 30~K, the fit leads to $\beta = 0.22\pm0.64$, and $M_{xy}$ reaches almost the value at high temperatures. Taking into account the large error bar of $\beta$ at 30~K, it seems that a combination of static fields and dynamic wipeout leads to the loss of signal intensity. For undoped LaFeAsO, the zero field NMR signal appears at about 11.6 MHz at low temperatures $\leq 8$ K \cite{MukudaJPSJ2009,MoroniPRB2017}. However, since we are still above the bulk \TN\ here, and the doping should lead to a reduction and distribution of magnetic moments, it may well be that the signal from ordered regions of the sample does not lie in the frequency range of Fig.~\ref{fig:TabhNQR}, or is too spread out to be observed. Indeed, a broadening in the spectra is clearly visible below \TS , and a strong broadening of the magnetic state NMR spectrum is also known for low Co and Ni concentrations in \BFA , especially just below $T_N$ \cite{NingPRB2014,DioguardiPRB2010}. Measurements with short separation times $t<10$~$\mu$s on large single crystals would be welcome to clarify this point.

\section{Spin-lattice relaxation rate and AC susceptibility}
\label{sec:T1andsusce}

\subsection{Spin-lattice relaxation rate measured by NMR ($H || ab$) and NQR (corresponding to $H || c$)}

Similar to the quadrupole parameters, the spin-lattice relaxation rate divided by temperature, \slrrt , is strongly doping dependent, and changes at the nematic transition \TS . The latter is because of the peculiar hyperfine coupling of the As nucleus due to the position of the As above and below the square of iron atoms (see Sec.~\ref{sec:NMR}), and due to the interplay of orbital and spin degrees of freedom.

\begin{figure}
\begin{center}
 \includegraphics[width=\columnwidth,clip]{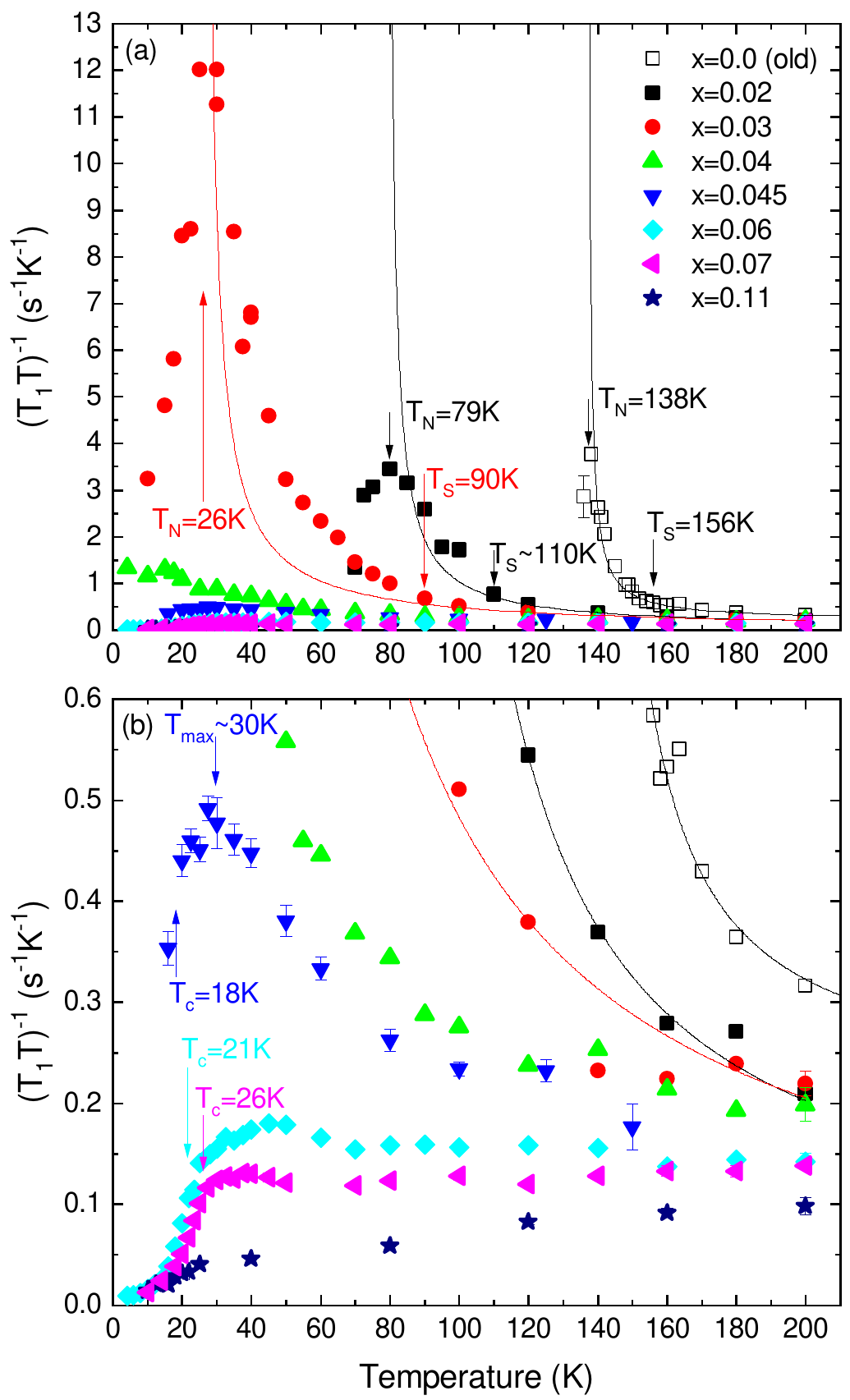}
 \caption{(Color online) The temperature dependence of the spin-lattice relaxation rate divided by temperature, \slrrt , in a magnetic field of $H||ab = 7$ T for \LaOFFeAsx\ with $x=0$ (open black squares), $x=0.02$ (black squares), $x=0.03$ (red dots), $x=0.04$ (green triangles), $x=0.045$ (blue down triangles), $x=0.06$ (cyan diamonds), $x=0.07$ (magenta left triangles), and $x=0.11$ (dark yellow hexagons). The data for $x=0$ is from Ref.~\onlinecite{HammerathPRB2013} and agrees well with data for single crystals \cite{OkPRB2018,FuPRL2012}. The upper panel highlights the data of the magnetic samples, whereas the lower panel is a zoom in to the data of the superconducting samples. \TN , \TS , \Tc , and \Tmax\ are marked with arrows. The solid lines are fits to a combined Curie Weiss and gap-like equation \cite{NingPRL2010}, and include data down to \TS\ only.}
\label{fig:T1T} 
\end{center}
\end{figure}

Fig.~\ref{fig:T1T} shows \slrrt\ for the different doping levels in a magnetic field of 7 T parallel to the $ab$ plane. The upper panel (Fig.~\ref{fig:T1T} (a)) shows the behavior of those samples which order magnetically, and the lower panel (Fig.~\ref{fig:T1T} (b)) is a zoom-in to the superconducting samples, whose relaxation rate is strongly reduced. In the temperature range of Fig.~\ref{fig:T1T}, \slrrt increases slowly with decreasing temperature, and then, at \TS , more strongly increases with decreasing $T$, even surpassing a Curie Weiss fit to the data above \TS . The temperature at which \slrrt\ increases more strongly than the Curie Weiss fit defines the structural phase transition temperature \cite{NingPRB2014}. The maximal value of \slrrt\ at \TN\ increases with doping for those samples that order long range magnetically, consistent with underdoped \BFCoAx \cite{DioguardiPRL2013}.

There is a drastic change in \slrrt\ between samples that order magnetically and those that are superconducting. For $x=0.03$, \slrrt\ reaches a maximum of about 12~s$^{-1}$K$^{-1}$ at $T_N = 26$~K, and for $x=0.04$, which does not order long range, \slrrt\ reaches only 1.34~s$^{-1}$K$^{-1}$ at 4.2~K, i.e. about one order of magnitude smaller. The sample with $x=0.045$ shows a broad maximum in \slrrt\ at $T_{\mathrm{max}} = 30$~K, but with an even more reduced \slrrt\ value of only 0.5~s$^{-1}$K$^{-1}$. Such a broad maximum has been observed already earlier and has been attributed to glassy freezing of spin fluctuations \cite{HammerathPRB2013}. At \Tmax , the inverse of the temperature dependent correlation time of the spin fluctuations, $\tau_c(T)$, equals the Larmor frequency of the nuclei, $\omega_L$, which leads to the maximum in \slrrt\ (for details the reader is referred to Ref.~\onlinecite{HammerathPRB2013} and references therein). Such a maximum can occur at a higher temperature than \TN\ for a long range magnetically ordered sample which is indeed the case here (compare $x=0.03$ and $x=0.045$). Our findings are consistent with earlier zero-field $\mu$SR and M\"o\ss{}bauer studies where indications of low-$T$ disordered magnetism were found in underdoped samples with 0.05~$\leq x \leq$~0.075, with internal fields being a factor of 20 smaller than in the magnetically ordered samples \cite{LuetkensNatMat2009}. Overall, this state is best described as a glass-like transition of the spin fluctuations with low temperature static disordered magnetism with reduced internal magnetic fields.

Alternatively, the maximum in such underdoped, superconducting samples has been taken as evidence for long range magnetic order with a greatly reduced moment of 0.006 $\mu_B$ \cite{YangSciChin2018} and 0.0072 $\mu_B$ \cite{NakaiPRB2012}, and ordering temperatures of $T_N = 58$ K and $T_N = 30$ K, respectively, compared to 0.6 $\mu_B$ and $T_N =$137~K in the parent compound \cite{QureshiPRB2010,GrafeNJP2009}. The height of the maximum in \slrrt\ in both of these samples is similar to what we observe for superconducting samples with $x=0.045$, and is much smaller than the maximum of our last long range ordered sample with $x=0.03$. Therefore, we believe that these samples from Ref.~\onlinecite{YangSciChin2018,NakaiPRB2012} do not order long range magnetically, but exhibit a progressive slowing down of spin fluctuations with a glass-like magnetic ordered state \cite{HammerathPRB2013}. Furthermore, we do not observe a structural phase transition in this doping range, which always appears above the magnetic phase transition (see below).

For even higher doping values, \slrrt\ is further reduced, and decreases strongly below \Tc\ due to the superconducting gap, following the typical power law behavior \cite{GrafePRL2008}. Finally, for $x=0.11$ no signature of spin fluctuations can be observed anymore in \slrrt , and it adopts the slope of the high temperature decrease of the other doping levels~\cite{HammerathPRB2013}, until \slrrt\ decreases in the superconducting state below \Tc .              

\begin{figure}
\begin{center}
 \includegraphics[width=\columnwidth,clip]{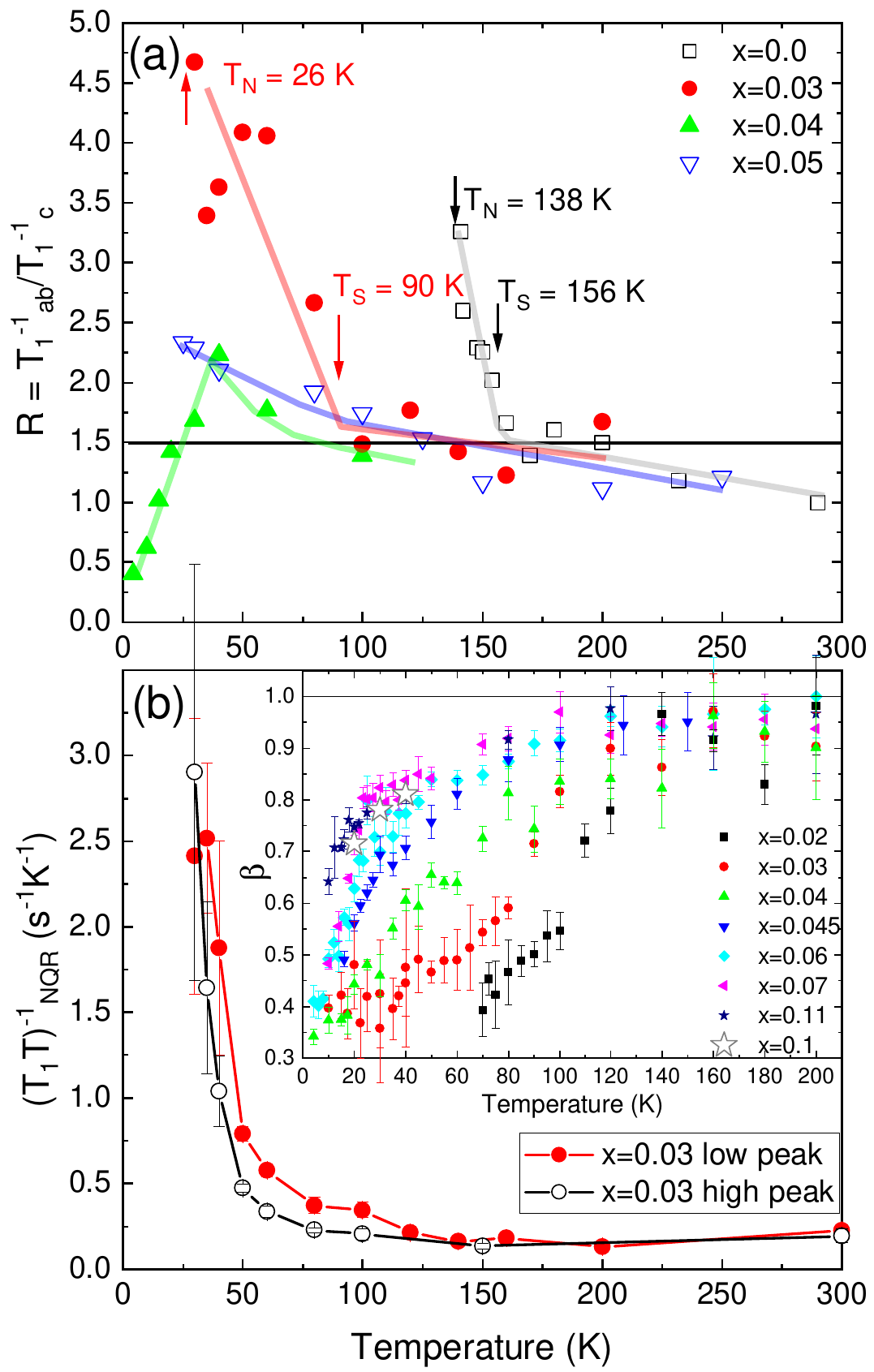}
 \caption{(Color online) (a) The temperature dependence of the ratio of the spin-lattice relaxation rates measured for different field orientation, \ratio . The open symbols indicate published data ($x=0$ (black open squares) and $x=0.05$ (blue open down triangles) \cite{LangPRL2010,HammerathPRB2013}, and the closed symbols are new data ($x=0.03$ (red dots), and $x=0.04$ (green triangles)). Lines are guides to the eye. (b) \slrrt\ versus $T$ measured by NQR on the low frequency peak and on the high frequency peak of $x=0.03$. Inset: The stretching parameter $\beta$ for different doping levels (all measured by NMR). A discontinuous drop at \TS\ is clearly visible for $x=0.02$ and $x=0.03$. Open stars indicate new fits including the stretching exponent of old data from Ref.~\onlinecite{GrafePRL2008}.}
\label{fig:T1ratio}
\end{center}
\end{figure}

For the undoped and underdoped samples, there is another way to determine the structural phase transition temperature \TS\ from \slrrt\ measurements that relies on the anisotropy of the hyperfine couplings, and therefore on the anisotropy of \slrr \cite{KitagawaJPSJ2009,SKitagawaPRB2010,NakaiPRB2012}. The ratio of the relaxation rates measured for $H||ab$ and $H||c$, i.e. \ratio , changes pronouncedly below \TS . Because we did not measure \slrrc\ in a magnetic field, we took the relaxation rate measured by NQR which is equivalent to \slrrc . The ratio, $R$, is shown in Fig.~\ref{fig:T1ratio} (a), and \slrrt measured by NQR on the two peaks for $x=0.03$ in Fig.~\ref{fig:T1ratio} (b). Due to the peculiar hyperfine coupling of the \as , $R$ is about 1.5 at higher temperatures, and increases strongly at the structural phase transition temperature. The so determined transition temperatures, \TS , coincide with those determined from \slrrab\ only. For samples, which do not show long range magnetic order at low temperatures ($x=0.04$ and $x=0.05$), the ratio increases smoothly above $R=1.5$, and reaches a maximal value of only $R=2.3$. For $x=0.04$, $R$ decreases again below 40~K and becomes smaller than 1.5 below 20~K. This indicates that the spin fluctuations are not stripe type anymore for this particular doping at low temperatures.

All samples show a distribution of spin-lattice relaxation rates similar to \BFCoAx\  which indicates dynamical inhomogeneity in the samples \cite{DioguardiPRL2013,DioguardiPRB2015}. This distribution can be quantified by a stretching exponent $\beta$ which is shown in the inset of Fig.~\ref{fig:T1ratio} (b) (see also Sec.~\ref{sec:NMR}). Interestingly, $\beta$ seems to drop discontinuously at \TS\ for $x=0.02$ and $x=0.03$ , and is therefore another indicator of the nematic transition. Note that also optimally and overdoped samples show a modest stretching of $\beta \approx 0.8$. In this case, the stretching exponent can be fixed to 1 as has been done in Ref.~\onlinecite{GrafePRL2008} without significantly reducing the quality of the fit.

Measurements of the spin-lattice relaxation rate on the two NQR peaks for $x=0.03$ show that the low frequency peak relaxes faster at low temperatures, and tends to saturate close to \TN , whereas the relaxation rate of the high frequency peak is slower, but seems to further increase down to the lowest measured temperature of 30 K (see Fig.~\ref{fig:T1ratio}). This behavior is consistent with the temperature dependent intensities of the two peaks, and is further evidence that the low-doping-like regions account for the magnetism in these samples.  

Overall, we could determine the nematic ordering temperature \TS\ from \slrrab\ and from the ratio $R$ which both coincide, and the magnetic ordering temperature \TN\ from the large maximum in \slrrab . Only samples which become long range magnetically ordered show a structural phase transition, and the gap between \TS\ and \TN\ increases with doping level. There is no evidence for bulk coexistence of magnetism and superconductivity in \LaOFFeAsx , as those samples which do show a small maximum in \slrrt\ do not order long range, but exhibit glassy freezing of spin fluctuations.

\subsection{AC susceptibility}
\label{sec:sus}

\begin{figure}
\begin{center}
 \includegraphics[width=\columnwidth,clip]{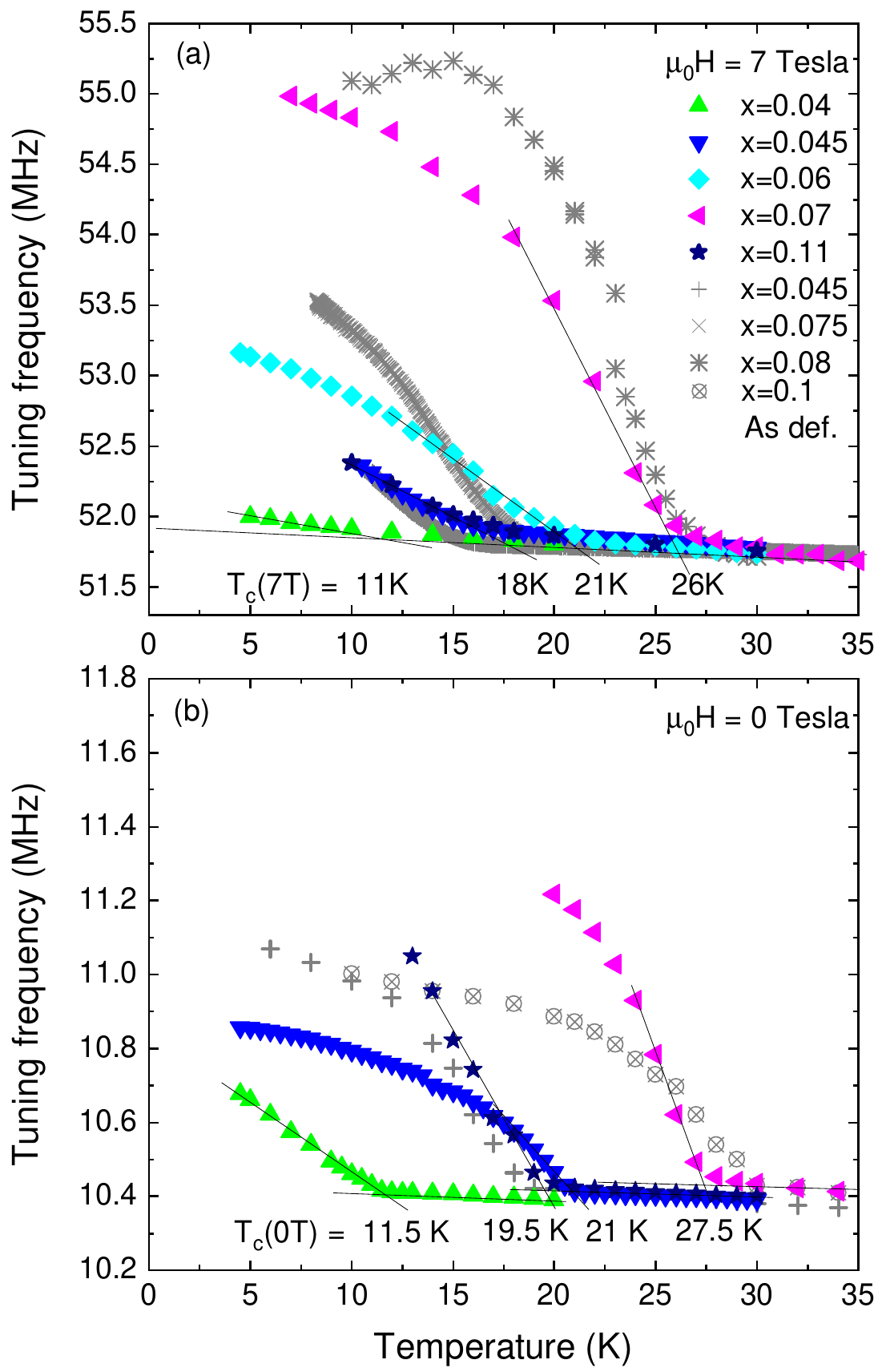}
 \caption{(Color online) The temperature dependence of the tuning frequency of the sample probe, which indicates the onset of superconductivity. (a) In 7 Tesla magnetic field. (b) In zero external magnetic field. The crossed grey symbols are older samples: $x=0.045$ from Refs.~\onlinecite{HammerathPRB2013,LangPRB2016}, $x=0.075$ from Ref.~\onlinecite{LuetkensNatMat2009}, $x=0.08$ from Ref.~\onlinecite{PrandoPRL2015}, and $x=0.1$ the "As deficient" sample from Refs.~\onlinecite{GrinenkoPRB2011,FuchsPRL2008,HammerathPRB2010}.}
\label{fig:acsus}
\end{center}
\end{figure}

The AC susceptibility of a sample can be measured by following the resonance frequency of the NMR or NQR sample probes resonant circuit depending on the temperature. This has been done in a magnetic field of 7~T (NMR) corresponding to a frequency of about 51.8~MHz, and in zero field (NQR) corresponding to a frequency of about 10.4~MHz. When a sample inside the coil of the resonant circuit becomes superconducting, the change in susceptibility and surface conductivity results in a change in the resonance frequency of the circuit. This way, one can check in situ the superconducting transition temperature of a sample. The AC susceptibility data is shown in Fig.~\ref{fig:acsus}. The \Tc\ values from our measurements coincide to within 1~K with those determined by the DC SQUID susceptibility. Note that AC measurements have not been done for all samples, and that this is only a semi-quantitative measurement, because the individual samples vary slightly and the coils which contain the samples differ between NMR and NQR measurements.

For the very underdoped sample with $x=0.04$, superconductivity is hardly visible in 7~T whereas for data taken at 0~T there is a significant change in the tuning frequency at \Tc . However, the \Tc 's themselves are very similar in 7~T and in zero field (11~K compared to 11.5~K, respectively).  This indicates that $x=0.04$ is not a bulk superconductor, and is compatible with a nanoscale separation with intrinsic superconductivity in the high-doping-like regions and superconductivity by proximity in the low-doping like regions \cite{LangPRB2016,PanarinaPRB2010}. The superconductivity by proximity is then suppressed at first by the magnetic field, and the relative change of the resonance frequency is much smaller due to the smaller volume fraction, but \Tc\ in the high-doping-like regions with intrinsic superconductivity is not much affected.  

The change in the resonance frequency and the superconducting transition temperature \Tc\ are maximal only for optimally doped samples around $x=0.07$, and the applied field has only little influence on both. For overdoped samples like $x=0.11$ the change in frequency  and the \Tc 's decrease again compared to optimal doping.

\section{Phase diagram and comparison to other samples}
\label{sec:phasedia}

\begin{figure}
\begin{center}
 \includegraphics[width=\columnwidth,clip]{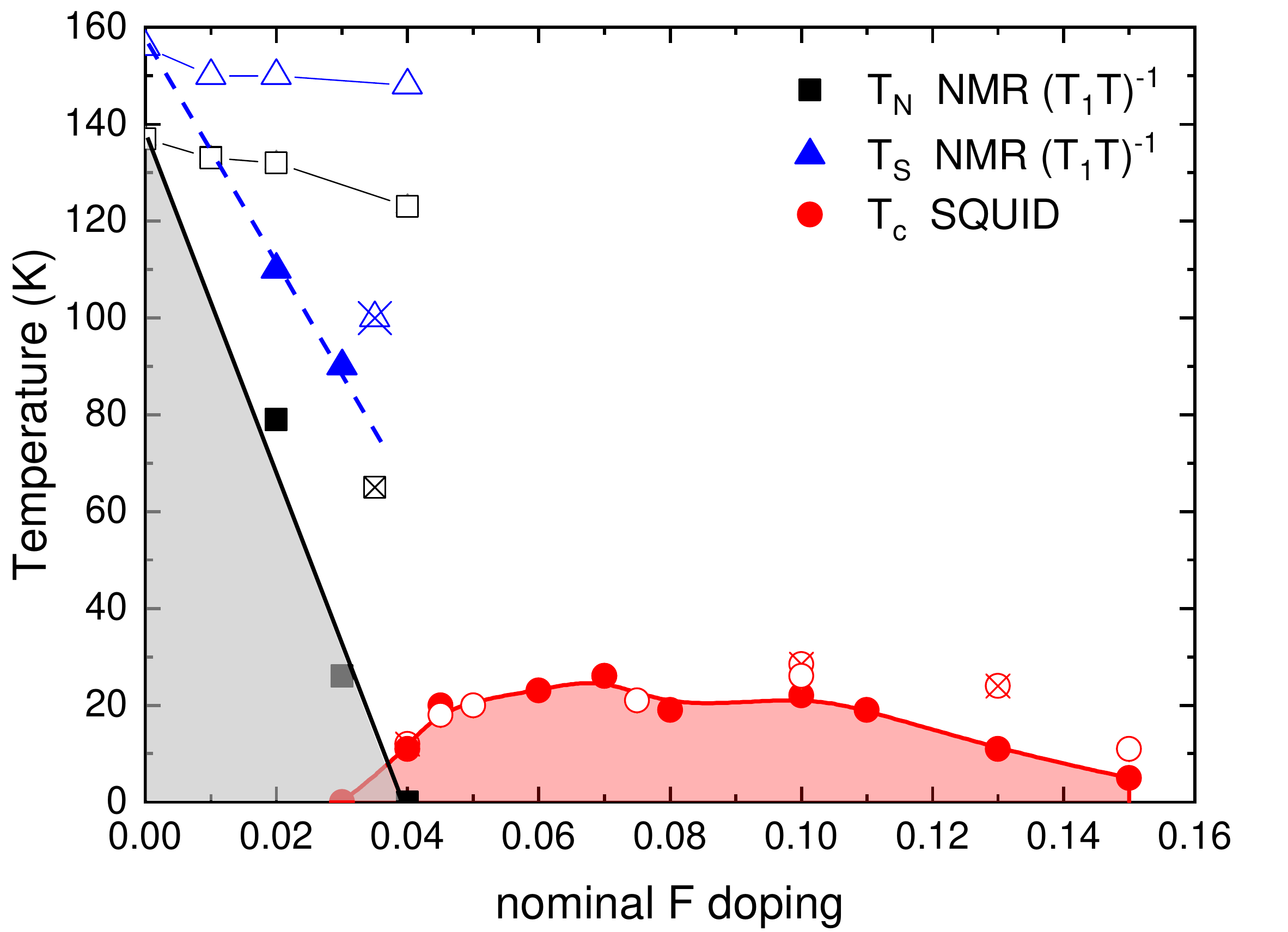}
 \caption{(Color online) The phase diagram of \LaOFFeAsx\ versus the nominal doping. The closed symbols are samples from this work, the open symbols are samples from Ref.~\onlinecite{LuetkensNatMat2009}, and the crossed open symbols are samples which were investigated over the years and which do not really fit into the phase diagram, for example $x=0.035$, $x=0.04$ from Refs.~\onlinecite{HammerathPRB2013,LangPRB2016}, a sample from Ref.~\onlinecite{PrandoPRL2015} which has a nominal content of $x=0.08$ and an EDX-determined content of $x_{\mathrm{EDX}}=0.13$, and the $x=0.1$ "As deficient" sample from Refs.~\onlinecite{GrinenkoPRB2011,FuchsPRL2008,HammerathPRB2010}. Note that \Tc\ of this sample is the 90 \% of $\varrho$ in the normal state value which does not indicate bulk superconductivity at that temperature. The lines are guides to the eye.}
\label{fig:phasedia}
\end{center}
\end{figure}

The extracted transition temperatures \TN , \TS , and \Tc , and the nominal F content can now be used to construct a phase diagram that is shown in Fig.~\ref{fig:phasedia}. For comparison, we include here the old samples from Ref.~\onlinecite{LuetkensNatMat2009} (open symbols), and some samples which have been published in Refs.~\onlinecite{HammerathPRB2013,LangPRB2016} ($x=0.035$, $x=0.04$, and $x=0.045$, crossed open symbols), in Ref.~\onlinecite{PrandoPRL2015} ($x_{\mathrm{nom}}=0.08$ and $x_{\mathrm{EDX}}=0.13$, crossed open symbol), and in Refs.~\onlinecite{GrinenkoPRB2011,FuchsPRL2008,HammerathPRB2010} ($x=0.1$ the "As deficient", crossed open symbol).

The new samples show an almost linear decrease of \TN\ and \TS\ which terminates at a nominal F content of about $x=0.04$. Such an almost linear dependence of \TN\ and \TS\ has been found recently in Co doped \LaOFeAs\ single crystals, too, and \TN\ terminates at a Co concentration of $x \approx 0.04$  \cite{HongArXiv2019,Lepuckiunpub2}.  Superconductivity also sets in at a fluorine content of about $x=0.04$, though the volume fraction of superconductivity is low, and the sample with $x=0.04$ is not a bulk superconductor. The maximal \Tc\ is reached around $x=0.07$, and stays almost constant up to $x=0.1$. For overdoped samples $x\geq 0.11$, \Tc\ and the superconducting volume fraction decrease again.      

\begin{figure}
\begin{center}
 \includegraphics[width=0.953\columnwidth,clip]{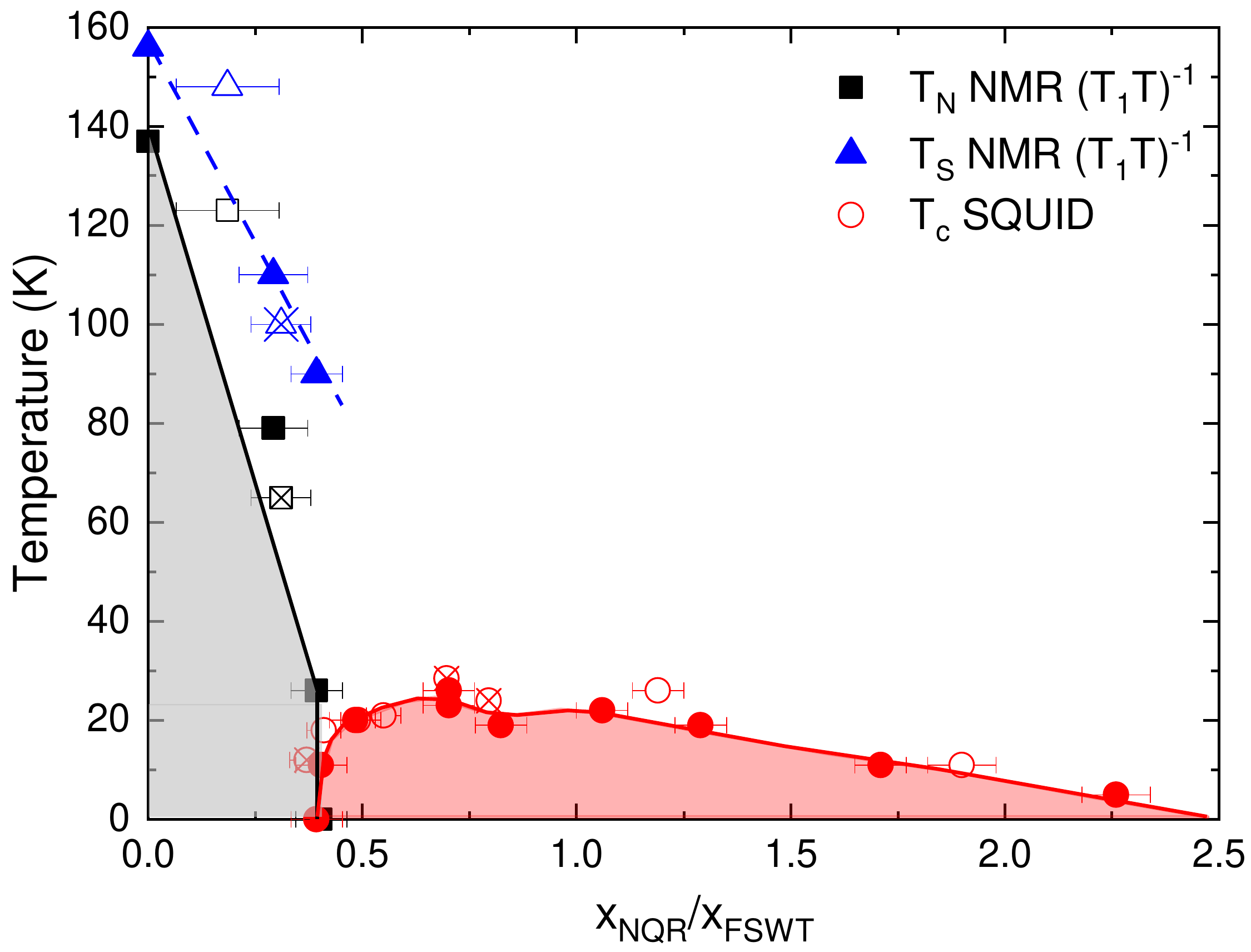}
 \caption{(Color online) The phase diagram of \LaOFFeAsx\ versus the doping level determined by NQR, $x_{\mathrm{NQR}}/x_{\mathrm{FSWT}}=w^H$. The closed symbols are samples from this work, the open symbols are samples from Ref.~\onlinecite{LuetkensNatMat2009}, and the crossed open symbols are samples which were investigated over the years. When the doping level is determined by NQR, all samples fit into this phase diagram within error bars. Note that there is not a one-to-one relation between $x_{\mathrm{NQR}}/x_{\mathrm{FSWT}}$ and the nominal F content, since the nominal F content varies. The lines are guides to the eye.}
\label{fig:phasediaNQR}
\end{center}
\end{figure}  

In comparison to the published phase diagram \cite{LuetkensNatMat2009}, differences occur mostly in the underdoped range from $x=0.01$ to $x=0.04$. In the new samples, \TN\ and \TS\ decrease much faster with doping than in the old samples. In order to reconcile the two phase diagrams, we plot in Fig.~\ref{fig:phasediaNQR} the transition temperatures versus the doping level that has been extracted from the high frequency weight $w^H=x_{\mathrm{NQR}}/x_{\mathrm{FSWT}}$ from the NQR spectra, and from the doping dependence of the NQR frequency for optimally- and overdoped samples. This procedure is described in detail in Ref.~\onlinecite{LangPRB2016}. $w^H$ is the spectral weight of the high frequency peak, and $x_{\mathrm{FSWT}}$ is the concentration, at which full spectral weight transfer (FSWT) occurs, i.e. when $w^H = 1$ and the low frequency peak has disappeared. As a result of this procedure, all samples can be unified within the error bars, and show a consistent doping dependence of the transition temperatures.  

The linear decrease of \TN\ and \TS\ for underdoped samples is preserved, and magnetism terminates close to the site percolation threshold for two dimensions which is 0.41 \cite{LangPRB2016,SykesPR1964,NewmanPRL2000}. This means, as soon as approximately 41~\% of the FeAs plane belong to the high doping like regions that contribute to $w^H$, magnetism disappears. Both the linear decrease and the threshold of 0.41 indicate that dilution of magnetic sites until the percolation threshold is indeed the reason for the disappearance of magnetism with doping, and that superconductivity has little to no influence. Note that the percolation threshold does not change much for single site percolation or clusters of 2-4~nm size \cite{SykesPR1964,NewmanPRL2000}. The cluster size of a few nm, i.e. the size of the high doping like regions, has been estimated from the superconducting coherence length in Ref.~\onlinecite{LangPRB2016}, and agrees with $\mu$SR results and NMR under pressure \cite{SannaPRB2009,SannaPRB2010,FujiwaraPRB2012}. Below the percolation threshold the size of the clusters is too small to harbor superconductivity, whereas above the percolation threshold, the low doping like regions are too small and disconnected to develop long range magnetic order. Beyond the percolation threshold, the clusters are connected and can therefore harbor superconductivity. The high doping like regions have potentially  polaronic character, where the conduction electrons are trapped by distortions of the surrounding lattice that are induced by the F dopants. With increasing F concentration, these regions do not overlap, but are connected and increase in size, so that the two well separated peaks in the NQR spectra are maintained, and the high frequency peak increases with the doping level.       

Moreover, certain anomalous values for higher dopings such as  $x_{\mathrm{EDX}}=0.13$ \cite{PrandoPRL2015}, and the $x=0.1$ "As deficient" \cite{GrinenkoPRB2011,FuchsPRL2008,HammerathPRB2010} could be captured. Note that also the spin-lattice relaxation rates and the AC susceptibilities are similar for similar NQR-determined doping levels (see Fig.~\ref{fig:acsus} for AC susceptibility, and Fig.~\ref{fig:oldT1T} in the APPENDIX for \slrrt ), suggesting that samples with similar NQR determined doping levels indeed have similar physical properties. In particular, the physical properties of these two samples could be possibly related to their doping level. According to the results presented here, they are indeed slightly underdoped samples with $x_{\mathrm{NQR}}=0.795$ ($x_{\mathrm{EDX}}=0.13$) and $x_{\mathrm{NQR}}=0.7$ ("As deficient" $x_{\mathrm{nom}}=0.1$). Disorder plays a central role to explain the high upper critical field values in the latter sample \cite{FuchsPRL2008} as well as the enhancement of the superfluid density, $\varrho_s$, while \Tc\ remains constant in the former sample \cite{PrandoPRL2015}. A natural source of disorder would be the low-doping-like regions. In fact, the enhancement of the upper critical field has also been observed in underdoped \LaOFFeAsx \cite{KohamaPRB2009}, and underdoped \SmOFFeAsx , and has been explained in the latter by strong pinning due to non-superconducting regions intermixed on the nanoscale with the superconducting phase \cite{PanarinaPRB2010}. The enhancement of superfluid density under pressure while \Tc\ remains constant as reported in Ref.~\onlinecite{PrandoPRL2015} may be simply related to the relatively flat \Tc\ in the doping range $x_{\mathrm{NQR}}=0.7$ to $x_{\mathrm{NQR}}=1.2$ (see Fig.~\ref{fig:phasediaNQR}), and the reduced superfluid density below $x_{\mathrm{NQR}}=1.2$ \cite{LuetkensNatMat2009}.

\section{Summary and Conclusion}

We have used \as\ NMR and NQR on more homogenized samples of \LaOFFeAsx\ to construct a phase diagram, where the nominal doping level, $x_{\mathrm{nom}}$, was better controlled. The transition temperatures, \TN , \TS , and \Tc , have been determined by \slrrt\ and AC and DC susceptibility. This phase diagram deviates from the published one \cite{LuetkensNatMat2009} especially for low F concentrations, however, if the doping level is determined from the NQR spectra, both phase diagrams can be reconciled. Frequency dependent intensity, \ssrr , and \slrrt\ measurements on underdoped samples at the boundary of magnetism and superconductivity support the picture of a nanoscale separation. The low-doping-like regions account for orthorhombicity and magnetism, and the high-doping-like regions accommodate superconductivity. The magnetism is suppressed by percolation, and superconductivity appears only above the percolation limit, when the majority of the high doping like regions are connected. There is no evidence for bulk coexistence of superconductivity and magnetism in \LaOFFeAsx . Underdoped samples, which show magnetic order, are not superconducting, and those which are superconducting, exhibit glassy freezing of spin fluctuations, but no long range magnetic order \cite{HammerathPRB2013}. The results in \LaOFFeAsx\ seem to be in contrast to \BFCoAx , where the majority of experimental results points towards a microscopic coexistence of superconductivity and magnetism for $x \leq 0.055$, nevertheless with inhomogeneous dynamical correlations \cite{BernhardPRB2012,DioguardiPRL2013,DioguardiPRB2015}. This in turn is reminding of the cuprates where inhomogeneities vary in strength between different families due to different disordered lattice potentials. In the cuprates, the origin of the inhomogeneities and of charge ordering in underdoped samples lies in the Mott physics and short-range antiferromagnetic correlations that originate from the parent compounds. \cite{KeimerNat2015} In the pnictides, selective Mott physics is more relevant for the hole doped side close to half-filling, where charge ordering indeed has been observed experimentally \cite{CivardiPRL2016,MoroniPRB2019}. For electron doping, Mott physics is not relevant \cite{deMediciPRL2014}, but Hund's coupling $J$ \cite{HauleNJP2009} can lead to correlations which can induce inhomogeneities and charge order as well \cite{IsidoriPRL2019,StadlerAnnaPhys2019}. Our experimental observation of a nanoscale charge separation in \LaOFFeAsx\ demonstrates that inhomogeneities or charge order is not limited to the hole-doped side, but also occurs in electron-doped iron pnictides. Further local probe measurements such as scanning tunneling microscopy are welcome to investigate this charge order. Single crystals, which are necessary for such measurements, are currently available only for Co doping, and preliminary NQR results show that a similar charge order occurs for Co doping as well \cite{Lepuckiunpub,WangJMMM2019,HongArXiv2019}.

\section{Acknowledgements}

The authors acknowledge G. Lang, A. U. B. Wolter and the late G. Behr for fruitful scientific discussions, and A. Lindner, C.G.F. Blum, M. Deutschmann, J. Hamman-Borrero, K. Leger, C. Malbrich, S. M\"uller-Litvanyi, R. Wachtel, and J. Werner for experimental and technical support. This work has been financially supported by the Deutsche Forschungsgemeinschaft (DFG) through the Priority Program GRK1621, through SPP1458 (Grants No. GR3330/2, No. BE1749/13, No. BU887/15), and through the Emmy Noether project No. WU595/3-1. S.A. acknowledges financial support from DFG via Grant No. AS523/4-1.

\section{Appendix}
\label{sec:appendix}

\begin{figure}
\begin{center}
 \includegraphics[width=\columnwidth,clip]{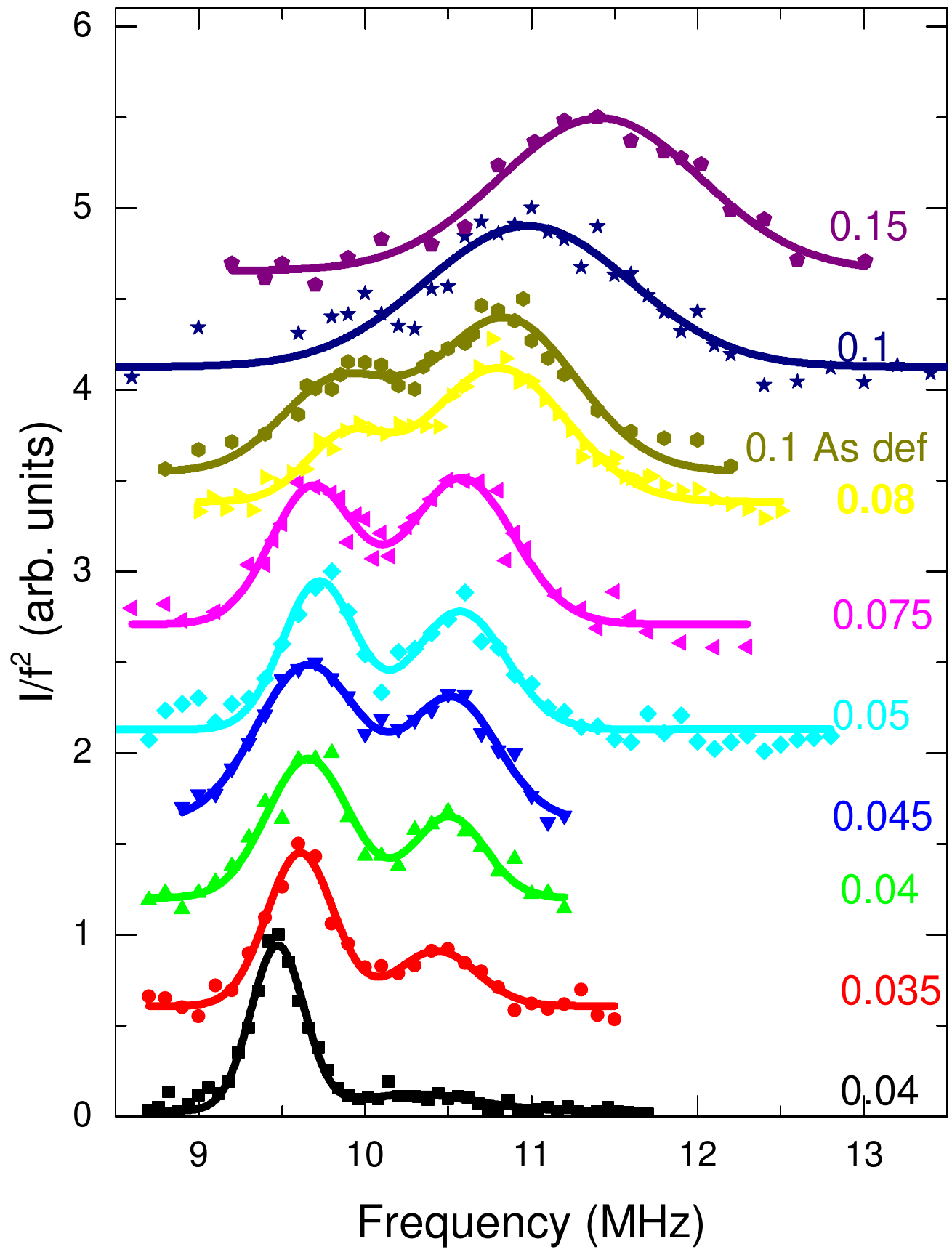}
 \caption{(Color online) The NQR spectra of other \LaOFFeAsx\ samples at room temperature. Only $x=0.1$ has been taken at 50 K, but the NQR spectrum of this doping level is not temperature dependent.}
\label{fig:oldNQR}
\end{center}
\end{figure}

The NQR spectra of the old samples look similar (see Fig.~\ref{fig:oldNQR} and compare to Fig.~\ref{fig:NQRspec}), however, the F content does not always fit. Most of this data is published in Ref.~\onlinecite{LangPRB2016}, except for $x=0.04$ (black squares, unpublished), $x=0.08$ is the sample from Ref.~\onlinecite{PrandoPRL2015}, but the NQR spectrum is unpublished, and $x=0.1_{\mathrm {Asd}}$ is the "As deficient" sample from Ref.~\onlinecite{GrinenkoPRB2011}.

\begin{figure}
\begin{center}
 \includegraphics[width=\columnwidth,clip]{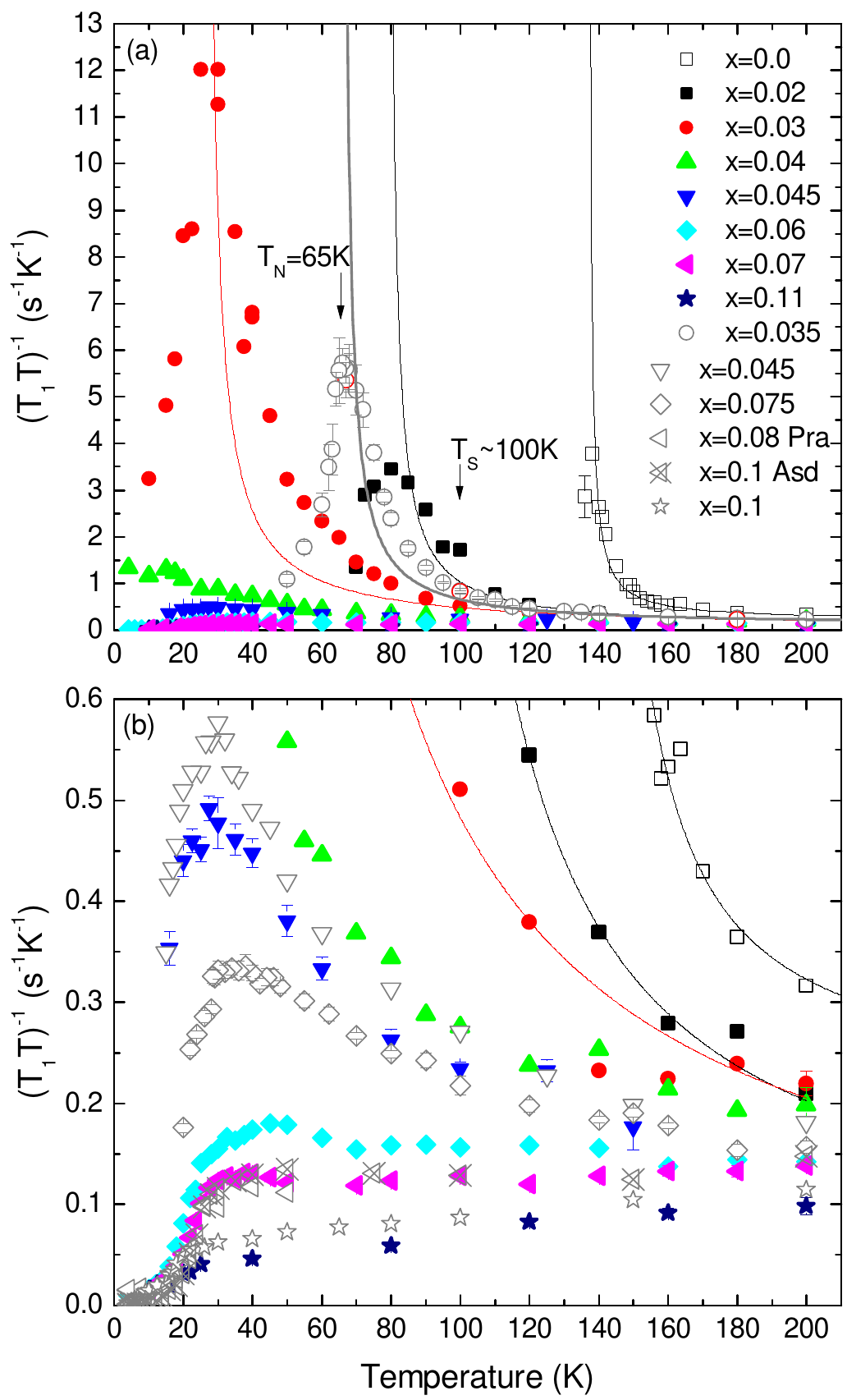}
 \caption{(Color online) \slrrt\ of other \LaOFFeAsx\ samples in comparison to the new samples.}
\label{fig:oldT1T}
\end{center}
\end{figure}

The temperature dependence of \slrrt\ in Fig.~\ref{fig:oldT1T} of the old samples fits to the new samples if the NQR doping is taken into account. Samples like the "As deficient" $x=0.1$ and the $x=0.08_{\mathrm Pra}$ sample with $x_{\mathrm{EDX}}=0.13$ show the same \slrrt\ temperature dependence as slightly underdoped samples. The $x = 0.035$ sample has a $T_N = 65$~K in between the new $x=0.02$ ($T_N = 79$~K) and $x=0.03$ ($T_N = 26$~K). Also the NQR determined doping level is in between: $x=0.02$ corresponds to $x_{\mathrm{NQR}}=0.292$, $x=0.035$ corresponds to $x_{\mathrm{NQR}}=0.31$, and $x=0.03$ corresponds to $x_{\mathrm{NQR}}=0.393$. The same is true for underdoped samples with $x=0.045$ (old and new), $x=0.05$, and $x=0.075$.

Finally, Tab.~\ref{tab:allsamples} summarizes all samples, their nominal F content, $x_{\mathrm{nom}}$, their NQR determined doping level, $x_{\mathrm{NQR}}$, their transition temperatures, \TN , \TS , and \Tc , and, regarding the old samples, their first publication reference.

\begin{table}
{
		\begin{tabular}{|c|c|c|c|c|c|c|c|c|c|c|c|}
		  \hline                                 \multicolumn{5}{|c|}{new samples} &  &  \multicolumn{6}{|c|}{old samples}  \\
			\hline \rule[-2ex]{0pt}{5.5ex} $x_{\mathrm{nom}}$ & $x_{\mathrm{NQR}}$ & \TN\ & \TS\ & \Tc\ & & $x_{\mathrm{nom}}$     & $x_{\mathrm{NQR}}$ & \TN\ & \TS\ & \Tc\ & Ref. \\ 
			\hline \rule[-2ex]{0pt}{5.5ex}           &           &      &      &      & &  0.0          &    0.0    & 137  &  156 & 0    & \onlinecite{LuetkensNatMat2009} \\
			\hline \rule[-2ex]{0pt}{5.5ex}   0.02    &  0.292    & 79   & 110  &  0   & &  0.04         &    0.185  & 123  &  148 & 0    & \onlinecite{LuetkensNatMat2009} \\
			\hline \rule[-2ex]{0pt}{5.5ex}   0.03    &  0.393    & 26   & 90   &  0   & &  0.035        &    0.31   & 65   &  100 & 0    & \onlinecite{HammerathPRB2013}   \\ 
			\hline \rule[-2ex]{0pt}{5.5ex}   0.04    &  0.404    & 0    & 0    & 11   & &  0.04         &    0.37   & 0    &  0   & 12   & \onlinecite{LangPRB2016}        \\ 
			\hline \rule[-2ex]{0pt}{5.5ex}   0.045   &  0.483    & 0    & 0    & 20   & &  0.045        &    0.41   & 0    &  0   & 18   & \onlinecite{HammerathPRB2013}   \\ 
			\hline \rule[-2ex]{0pt}{5.5ex}   0.06    &  0.702    & 0    & 0    & 23   & &  0.05         &    0.49   & 0    &  0   & 20   & \onlinecite{LuetkensNatMat2009} \\ 
			\hline \rule[-2ex]{0pt}{5.5ex}   0.07    &  0.702    & 0    & 0    & 26   & &  0.075        &    0.55   & 0    &  0   & 21   & \onlinecite{LuetkensNatMat2009} \\ 
			\hline \rule[-2ex]{0pt}{5.5ex}   0.08    &  0.823    & 0    & 0    & 19   & &  0.01$_{\mathrm{Asd}}$ &    0.7    & 0    &  0   & 29   & \onlinecite{FuchsPRL2008}       \\
			\hline \rule[-2ex]{0pt}{5.5ex}   0.1     &  1.06     & 0    & 0    & 22   & &  0.08$_{\mathrm{Pra}}$ &    0.795  & 0    &  0   & 24   & \onlinecite{PrandoPRL2015}      \\ 
			\hline \rule[-2ex]{0pt}{5.5ex}   0.11    &  1.29     & 0    & 0    & 19   & &  0.1          &    1.19   & 0    &  0   & 26   & \onlinecite{LuetkensNatMat2009} \\ 
			\hline \rule[-2ex]{0pt}{5.5ex}   0.13    &  1.71     & 0    & 0    & 11   & &  0.15         &    1.9    & 0    &  0   & 11   & \onlinecite{LuetkensNatMat2009} \\ 
			\hline \rule[-2ex]{0pt}{5.5ex}   0.15    &  2.26     & 0    & 0    & 5    & &               &           &      &      &      &                                 \\
			\hline 
		\end{tabular}}
		\caption{Left side: new samples and their nominal doping level, $x_{\mathrm{nom}}$, NQR determined doping levels, $x_{\mathrm{NQR}}$, \TN , \TS , and \Tc . Right side: old samples and their nominal doping level, NQR determined doping levels, \TN , \TS , and \Tc , and their first publication reference.}
		\label{tab:allsamples}
	\end{table}

\end{document}